\documentclass[useAMS,usenatbib]{mn2e}
\usepackage{graphicx,rotate,url,mathptmx,lscape,times}
\def\gtsim{\mathrel{\hbox{\rlap{\hbox{\lower4pt\hbox{$\sim$}}}\hbox{$>$}}}}
\def\lesssim{\mathrel{\hbox{\rlap{\hbox{\lower4pt\hbox{$\sim$}}}\hbox{$<$}}}}

\def\h0{\hbox{{\rm H}$^0$}}
\DeclareMathAlphabet{\vib}{OML}{cmm}{m}{it}

\title[Uncertainties in He~I Emissivities]{Uncertainties in Theoretical He~I Emissivities: H~II Regions, Primordial Abundance, and Cosmological Recombination}

\author[R.L. Porter et al.]
       {\parbox[]{6.0in}
        {R.L. Porter$^{1,2}$\thanks{E-mail: rporter@pa.uky.edu},
        G.J. Ferland$^{1,2}$,
        K.B. MacAdam$^{1}$,
        P.J. Storey$^{3}$\\
        \footnotesize
        $^1$Dept. of Physics \& Astronomy, University of Kentucky, Lexington, KY 40506, USA\\
        $^2$Institute of Astronomy, University of Cambridge, Madingley
        Road, Cambridge CB3 0HA, UK\\
        $^3$Dept. of Physics \& Astronomy, University College London, Gower Street, London WC1E 6BT, UK}}

\date{
      Received }

\pagerange{\pageref{firstpage}--\pageref{lastpage}}
\pubyear{}

\begin{document}

\maketitle

\label{firstpage}

\begin{abstract}

\noindent
A number of recent works in astronomy and cosmology have relied upon theoretical He~I emissivities, but we know of no effort to quantify the uncertainties in the atomic data.  We analyze and assign uncertainties to all relevant atomic data, perform Monte Carlo analyses, and report standard deviations in the line emissivities.  We consider two sets of errors, which we call ``optimistic'' and ``pessimistic.''  We also consider three different conditions, corresponding to prototypical Galactic and  extragalactic H~II regions and the epoch of cosmological recombination.  In the extragalactic H~II case, the errors we obtain are comparable to or larger than the errors in some recent $Y_p$ calculations, including those derived from CMB observations.  We demonstrate a systematic effect on primordial abundance calculations; this effect cannot be reduced by observing a large number of objects.  In the cosmological recombination case, the errors are comparable to many of the effects considered in recent calculations.

\end{abstract}


\section{Introduction}
\label{intro}
The need for accurate theoretical helium emission predictions is present in many aspects of astronomy and cosmology.  These include estimates of the primordial helium abundance, calculations of the cosmological recombination history, and, most recently, estimates of the time variation of the Higgs vacuum expectation value (Ga\ss ner, Lesch, \& Arenh\"ovel 2008).  However, these works may involve underestimates of the uncertainties in He~I emissivities.  


Standard Big Bang Nucleosynthesis (SBBN) yields are generally a function of a single parameter, the primordial baryon-to-photon ratio, $\eta$.  An important test of SBBN is that, for a single value of this parameter, the predicted light nuclei abundances agree with the abundances deduced from observations.  This is usually expressed via a concordance diagram (see Kirkman et al. 2003).  CMB observations have tightly constrained $\eta$ (Spergel et al. 2007).  However, the primordial He abundance, $Y_p$, deduced from observations sometimes disagrees with the SBBN yield.  This is troubling because He is the most abundant of these nuclei.  This is an area of active research (Fukugita \& Kawasaki 2006; Peimbert, Luridiana, \& Peimbert 2007; Izotov, Thuan, \& Stasi\'nska 2007, hereafter FK06, PLP07, and ITS07, respectively).  The usual method for deducing $Y_p$ is to construct a relation between metallicity (typically taken as the oxygen and/or nitrogen abundances) and the helium abundance in selected extragalactic objects.  The $dY/dZ$ relation is then extrapolated to zero metallicity (at which point all nuclei abundances should be equal to their primordial values).  For a recent review of the subject, see Steigman (2007). Many authors have discussed the errors involved in the $dY/dZ$ relation analyses (Skillman et al. 1998; Peimbert et al. 2003; Porter et al. 2007).
 
The study of the cosmological recombination spectrum is also currently an area of active research (Kholupenko, Ivanchik, \& Varshalovich 2007; Chluba \& Sunyaev 2008; Wong, Moss, \& Scott 2008).  The majority of effects considered in recent years modify the electron fraction by $<0.1\%$ (see Switzer \& Hirata 2008).  

In this Letter we seek to address the errors in  theoretical He~I emissivities.  We have previously calculated He~I emissivities with two different codes (Bauman et al. 2005; Porter et al. 2005, 2007).  In the low-density limit Bauman et al. (2005) compares the results of these two calculations.  We note that the differences listed there are not meant to be taken as errors.
Here we analyze available information and attempt to quantify the accuracy of all atomic data involved in our model atom (Porter et al. 2005).  We then quantify the uncertainties in emissivities via two methods: 1) a rigorous propagation of error in the low-density limit; and 2) random perturbations in a Monte-Carlo analysis.  We consider three sets of conditions typical of Galactic H~II regions, extragalactic H~II regions, and the epoch of cosmological recombination.  Finally, we demonstrate that, while our analysis involves only statistical errors, the errors will yield systematic effects in primordial helium calculations.  The most recent calculations of $Y_p$ (PLP07, ITS07) claim uncertainties $\lesssim1\%$, including estimates of the atomic data uncertainties.  We demonstrate that atomic data uncertainties alone introduce a systematic error of similar size.  
 
\section{Atomic Data}

The physical processes discussed here are largely described in Osterbrock \& Ferland (2006).  We group the relevant atomic data by physical process and present them in order of generally increasing uncertainty as follows: bound-bound radiative transitions, free-bound radiative transitions, and collisions.  Additional sources of uncertainty are also considered in Section~\ref{other_uncertainties} below.   Level energies are by far the most accurate data involved, and we neglect these uncertainties entirely.  

Because some atomic data uncertainties will have a nearly linear effect on corresponding emissivity uncertainties, it is critically important what uncertainties we assign to these data.  In an attempt to mitigate this, we consider both pessimistic and optimistic cases, which are meant to respectively represent the minimum and maximum uncertainty in each datum.  In all cases we allow only one significant figure in our uncertainties, with optimistic cases rounded down and pessimistic cases rounded up.

\begin{table}
\centering
\caption{Assumed Uncertainties in Helium Atomic Data.}
\begin{tabular}{lrr}
\hline
Conditions &  Optimistic & Pessimistic \\
\hline
Rad. Recomb. Coefficients (Direct)\\
\hline
$n>=5$ and $L>3$ & 0\% & 0.1\%\\
$n>=5$ and $L<=3$ & 0.01-0.7\% & $\leq4$\%\\
$n<5$ & 0.01-0.7\% & $\leq4$\%\\
\hline 
E1 Transition Probabilities\\
\hline
$n_u$,$n_l<10$ and $L<7$ & 0.01\% & 0.2\% \\
$n_u$,$n_l<10$ and $L>=7$ & 0\% & 0.01\%\\
$n_u>10$, $n_l<5$ and $L_l<=2$ & 0.02\% & 0.2\%\\
$n_u>10$, $L_u\geq2$ and $L_l\geq2$ & 0.6\% & 4\%\\
$n_u>10$, $n_l<10$ other & 1\% & 7\%\\
$n_u,n_l>10$ & 10\% & 10\% \\
\hline 
Other Transition Probabilities\\
\hline
$2p\,{}^{3}$P$_1 - 1s\,{}^{1}$S & 1\% & 5\% \\
$2p\,{}^{3}$P$_2 - 1s\,{}^{1}$S & 1\% & 1\% \\
$2s\,{}^{3}$S - $1s\,{}^{1}$S ($2\nu$) & 10\% & 30\% \\
$2s\,{}^{3}$S - $1s\,{}^{1}$S (M1) & 1\% & 20\% \\
$2s\,{}^{1}$S - $1s\,{}^{1}$S & 1\% & 5\% \\
all others & 1\% & 1\% \\
\hline
Collisional Deexcitation\\
\hline
$n_u<=5$ and $n_l<=2$ & 10\% & 30\%\\
$\Delta n =0$ & 20\% & 30\%\\
otherwise & 20\% & 30\% \\
\hline
\end{tabular}
\label{tab:uncertainties}
\end{table}

\subsection{Transition probabilities}

Electric-dipole (E1) transition probabilities between low-lying levels are extremely accurate.  Drake (1996) calculated values for $n \leq 10$ and $L \leq 7$.  We will refer to these levels as the ``Drake set.'' He claims ``essentially exact results for the entire singly-excited spectrum of helium.''  For a small subset of these transitions, Argenti \& Moccia (2008) compare their independent calculations with the corresponding Drake oscillator strengths; these agree to 5 or 6 significant figures.  Argenti \& Moccia also discuss discrepancies between their results using velocity, length, and acceleration gauges and find that the acceleration gauge occasionally produces results that differ by one part in $10,000$.  We therefore conservatively adopt the still negligible uncertainty of $0.01\%$ for the Drake transition probabilities.   For the pessimistic case, we take $0.2\%$, an estimate of the higher order relativistic corrections given by Drake \& Morton (2007).

Next we consider E1 transitions with upper level beyond the Drake set and lower level within the Drake set. These transitions have 
$n_u \geq 11$ or $L_u \geq 8$ and $n_l \leq 10$ or $L_l \leq 7$.  We use several different algorithms for these transitions.  
For $n_l \leq 5$ and $L_l \leq 2$, we extrapolate Drake results as in Burgess \& Seaton (1960).  
For transitions with both upper and lower $L \geq 2$, we use hydrogenic rates. For all other transitions with lower level within the Drake set, we use Drake's (1996) semi-classical algorithm for calculating radial integrals from quantum defects.  These algorithms agree with the tabulated Drake values to better than $0.05\%$,  $4\%$, and  $7\%$, respectively.  On average, these respective algorithms agree to $0.02\%$,  $0.6\%$, and  $1.0\%$.  We adopt these maximum and average differences as the uncertainties of transition probabilities calculated with these algorithms in the pessimistic and optimistic cases, respectively.  Remaining E1 transitions have both upper and lower levels beyond the Drake set.  Some of these rates (for example, with large $n$ and small $\Delta n$) are expected to be significantly less accurate than the low-lying transitions calculated by Drake.   We find, however, that because of the small contributions these levels have to the effective recombination coefficients of low-lying levels, the uncertainties in these transition probabilities are unlikely to contribute more than $0.01\%$ to the uncertainties of low-lying emissivities.  We assume $10\%$ uncertainties for all of these transitions.     

Forbidden transitions are generally less accurate than their dipole-allowed counterparts.  For the intercombination line $2p\,{}^{3}$P$_1 - 1s\,{}^{1}$S we take the $1\%$ band of theoretical values for the optimistic case and the roughly $5\%$ experimental uncertainty for the pessimistic case (Dall et al. 2008).  For the much slower $2p\,{}^{3}$P$_2 - 1s\,{}^{1}$S transition we take the transition probability from \L ach \& Pachucki (2001) and assume an uncertainty of $1\%$ (although this transition should be completely negligible in all conditions).  For the two-photon transition $2s\,{}^{1}$S - $1s\,{}^{1}$S, we take $1\%$ and $5\%$ for the respective optimistic and pessimistic cases.  The former is roughly the dispersion in theoretical values (Drake 1996; Jacobs 1971; Derevianko \& Johnson 1997), while the latter is taken from the uncertainty in the experimental lifetime of $2s\,{}^{1}$S given in Table~3 of Derevianko \& Johnson (1997).  The two-photon E1 transition $2s\,{}^{3}$S - $1s\,{}^{1}$S is assigned uncertainties $10\%$ and $30\%$, based upon the standard deviation and spread in Table~4 of Derevianko \& Johnson (1997).  The corresponding M1 transition is assigned $1\%$ and $20\%$ based upon a discussion by \L ach \& Pachucki (2001).  While more exotic transitions may be important in the cosmological recombination context, their uncertainties should contribute negligibly to the uncertainties in our solutions.  We assign $1\%$ errors to these transitions.

\subsection{Free-bound radiative rates}

Photoionization cross-sections (used to calculate recombination coefficients via the usual Milne relation) are calculated in a variety of ways which we will not repeat here.  Uncertainties in the optimistic case are taken as follows.  For $n \leq 4$, the threshold cross-section uncertainties are taken from the difference between the \textit{ab initio} and ``extrapolated'' results presented in Table~1 of Hummer \& Storey (1998).  For levels with larger $n$ and $L\leq2$, we simply use the same uncertainty that we assumed for $4{^{2S+1}}L$.  For levels with $L>2$, we take the (already quite small) uncertainty assumed at $L=2$.  For simplicity, we assume in the optimistic case that the energy-dependence of cross-sections above threshold is exactly known.  In this approximation, uncertainties in threshold cross-sections are equivalent to uncertainties in recombination coefficients, which we report in Table~\ref{tab:uncertainties}.

In the pessimistic case, however, we apply uncertainties to recombination coefficients directly (rather than indirectly via the threshold photoionization cross-sections).  For levels with $n\leq10$, we set the uncertainty equal to the difference between the two codes discussed in Bauman et al. (2005).  These differences reach as much as $4\%$ at $T_e=10^4$~K.  We apply the results at $n=10$ to all higher levels.  Dielectronic recombination onto He$^+$ forming He$^0$ becomes important only at $T_e>50,000$~K, and in the cases we consider here it can be neglected entirely.


\subsection{Collision rates}

Collision rates are generally the least accurate data in the models (although they do not necessarily contribute the most to the total uncertainty in emissivities).  Collisions between levels with $\Delta n \geq 1$ are most efficiently driven by electron impact.  Collisional deexcitation (and excitation) from levels with $n_u \leq 5$ and $n_l\leq2$ are calculated using the close-coupling R-matrix results of Bray et al. (2000), and from clues therein we take $10\%$ and $30\%$ to be the optimistic and pessimistic uncertainties.  For other transitions with $\Delta n \geq 1$ we employ various other algorithms and assume optimistic and pessimistic uncertainties of $20\%$ and $30\%$, respectively. 

Angular-momentum changing (or Stark) collisions are most efficiently driven by slow-moving particles.  In practice, these collisions are induced by protons.  For non-degenerate transitions, defined for this purpose as transitions with $\Delta n =0$ and $L \leq 2$ and $\Delta l = \pm1$, we use Seaton (1962).  In the energy-degenerate case we use the theory of Vrinceanu \& Flannery (2001), which naturally treats $\Delta l>1$ transitions.  While these ``l-mixing'' collisions are important for driving highly excited states into statistical equilibrium, they are considerably less important for small $n$.  We assume $20\%$ and $30\%$ for the optimistic and pessimistic uncertainties.

Recent results suggest large uncertainties in Rydberg level collisional ionization rates.  Nagesha \& MacAdam  (2003) found  experimental results that differed with respect to theoretical values by more than an order of magnitude.  Vrinceanu (2005) has suggested additional physical processes are at work.  Deutsch et al. (2006) extended a separate theory of these collisions and also found results much less than the experimental results.  Here we investigate the effects of changing collisional ionization rates by large factors.  Disabling collisional ionization entirely at $n_e = 10^6$~cm$^{-3}$ causes line emissivities to increase by $\approx1\%$ or less.  Multiplying collisional ionization rates by $10$ causes emissivities to decrease by $\approx1\%$ or less.  The three cases we will consider in Section~\ref{sec:montecarlo} are each significantly less dense than this test case, so the effects of collisional ionization are even less important.  We also note that collisional ionization should affect the Rydberg levels of He$^0$ and H$^0$ in similar ways.  This means that any uncertainty in He~I emissivities due to collisional ionization will be counteracted, at least to some degree, by uncertainty in H~I emissivities when He~I line fluxes are measured relative to H~I lines.  We conclude that collisional ionization uncertainties are completely negligible.  
		
\subsection{Other uncertainties}
\label{other_uncertainties}

The effect of the mixing of the singlet and triplet levels was considered by Bauman et al. (2005).  The effect was found to be negligible for multiplet emissivities in the low-density limit.  This is expected due to the principle of spectroscopic stability (Condon \& Shortley 1991).  That result may not hold at finite densities.  To investigate this question, we use a mixing algorithm based on the method outlined by Drake (1996) and find effects that are negligible in comparison with the assumed recombination coefficient errors.  We note that Rubi\~{n}o-Martin, Chluba, \& Sunyaev (2008) find interesting fine-structure absorption features in their $J$-resolved calculation of the cosmological He~I recombination spectrum.

The problems involved in modelling an (in principle) infinite set of levels with a finite system has been discussed at length by Porter et al. (2005) and Bauman et al. (2005).  Because we need to run many models for the Monte Carlo analysis, we use a smaller, less computationally intensive model atom than was used in our previous works.  We resolve all $nLS$ terms with $n\leq40$.  While this smaller model yields emissivities that differ by nearly a percent from our larger model for some low-lying transitions, we performed a small Monte Carlo calculation with the larger model and found the dispersion in results comparable to the dispersion with the smaller atom.  

\section{Low-Density, Case B Limit}

In the collisionless,
Case-B scenario, emission coefficients can be calculated without the inversion of a rate matrix.
Rather, level populations are calculated by considering decay and cascade probabilities 
$P_{ul}$ and $C_{ul}$ (defined following Robbins 1968).  
We use the uncertainties assigned in Table~\ref{tab:uncertainties} and perform a rigorous propagation of error,
neglecting any possible covariances not detailed above.
The uncertainty in the emission coefficient $4 \pi J / n_e n_{He^+}$ is given by

\begin{equation}
\sigma_J = h\nu_{ul} \sqrt{ (\sigma_{\alpha_{u}^{eff}} P_{ul})^2 + (\sigma_{P_{ul}} {\alpha_{u}^{eff}})^2}
\label{sigma_j}
\end{equation}

\noindent where $\alpha_{u}^{eff}$ is the \textit{effective} recombination coefficient (including indirect recombinations
from higher bound levels).

As mentioned above, E1 transition probabilities, and therefore decay probabilities, are well-known for transitions between
levels with low $n$.  Thus equation \ref{sigma_j} reduces to a linear relationship between $\sigma_J$
and $\sigma_{\alpha_{u}^{eff}}$, and the uncertainties in dipole emission coefficients (in this collisionless, Case-B scenario)
are entirely due to uncertainties in the effective recombination of the initial, upper level.  This low-density analysis serves mostly as a check on our Monte Carlo results.  
Our low-density results are similar to the results of the  `extragalactic' Monte Carlo models discussed below.

\section{Monte Carlo Calculations}
\label{sec:montecarlo}

We define our errors as one-standard-deviation uncertainties.  We disallow errors greater than three standard deviations from the nominal value because large deviations will inevitably produce unphysical trends with respect to quantum number.  For the random number generator (RNG) we use an implementation of the very high periodicity Mersenne Twister (Matsumoto \& Nishimura 1998), and we uniquely seed each calculation. We perform 1000 independent calculations for each model discussed below and for each set of error conditions in Table~\ref{tab:uncertainties}.    

Our three cases are defined as follows: 1) `Galactic' - a case B (Baker \& Menzel 1938) calculation with $T_e = 10,000$~K and $n_e = 10^4$~cm$^{-3}$ (similar to Models I and II of Porter et al. 2007); 2) `Extragalactic' - same as the Galactic case but with $T_e = 15,000$~K and $n_e = 10^2$~cm$^{-3}$; and 3) `Cosmological' - the He~I recombination epoch.  The actual recombination history is known to be both non-time-steady and non-LTE (Seager, Sasselov, \& Scott 2000).  Here we consider the much less computationally intensive time-steady case.  This model represents snapshots of recombining He$^+$ in an extremely intense radiation field with cosmological parameters from Switzer \& Hirata (2008a) and Sobolev optical depths.    

\section{Results and Discussion}

In Table~\ref{table:random}, we present the standard deviation of the Monte Carlo emissivities for a number of emission lines.  Note that one pair of lines is from the same upper level, $4p\,{}^{3}$P.  We obtain the necessary result that these lines have identical uncertainties.

\begin{table}
\centering
\caption{Standard deviations ($\%$) of emissivities in our Monte Carlo analyis.  Some trailing zeros are present for ease of comparison.}
\begin{tabular}{rrrrrrr}
\hline
  & & & \multicolumn{2}{|c|}{Galactic} & \multicolumn{2}{|c|}{Extragalactic} \\
$\lambda$ & $nl\,{}^{2S+1}\! L $ &  $nl\,{}^{2S+1}\! L$	& Opt. & Pess. & Opt. & Pess.   \\
\hline
 3965 & $4p\,{}^{1}$P & $2s\,{}^{1}$S & 0.3 &  0.8 & 0.07 & 0.2 \\
 4471 & $4d\,{}^{3}$D & $2p\,{}^{3}$P & 0.4 &  1.0 & 0.07 & 0.4 \\
 5876 & $3d\,{}^{3}$D & $2p\,{}^{3}$P & 1.0 &  4.0 & 0.20 & 0.8 \\
 6678 & $3d\,{}^{1}$D & $2p\,{}^{1}$P & 0.5 &  2.0 & 0.06 & 0.3 \\
 7065 & $3s\,{}^{3}$S & $2p\,{}^{3}$P & 6.0 & 20.0 & 0.80 & 3.0 \\
10830 & $2p\,{}^{3}$P & $2s\,{}^{3}$S & 8.0 & 30.0 & 2.00 & 9.0 \\
11970 & $5d\,{}^{3}$D & $3p\,{}^{3}$P & 0.2 &  0.9 & 0.05 & 0.4 \\
12530 & $4p\,{}^{3}$P & $3s\,{}^{3}$S & 0.8 &  2.0 & 0.09 & 0.4 \\
19540 & $4p\,{}^{3}$P & $3d\,{}^{3}$D & 0.8 &  2.0 & 0.09 & 0.4 \\
20580 & $2p\,{}^{1}$P & $2s\,{}^{1}$S & 5.0 & 10.0 & 0.30 & 1.0 \\
\hline
\end{tabular}
\label{table:random}
\end{table}



\begin{figure}
\protect\resizebox{\columnwidth}{!}
{\includegraphics{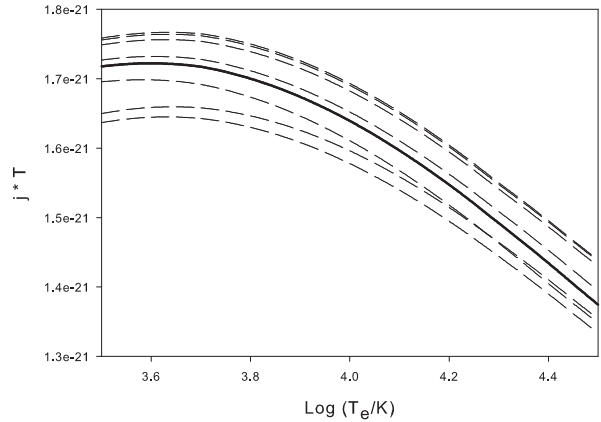}}
\caption{The emissivity of $\lambda5876$ (multiplied by $T_e$ for easier viewing) for several random sets (dashed curves) of perturbed data versus electron temperature and the unperturbed result (bold solid curve).  }
\label{fig:tweak}
\end{figure}

In Figure~\ref{fig:tweak} we investigate the effects of the present statistical analyses on $Y_p$ estimates.  Each line represents a single full set of random perturbations (corresponding to a particular seeding of the RNG).  For each set we calculate $j(\lambda 5876)$ for the range of temperatures shown on the x-axis.  To first order all curves are linear offsets of the unperturbed line.  This implies that statistical errors in He~I emissivities would affect an ensemble of He abundances all in the same sense and to roughly the same degree, yielding a systematic offset in any $dY/dZ$ relation.

\begin{figure}
\protect\resizebox{\columnwidth}{!}
{\includegraphics{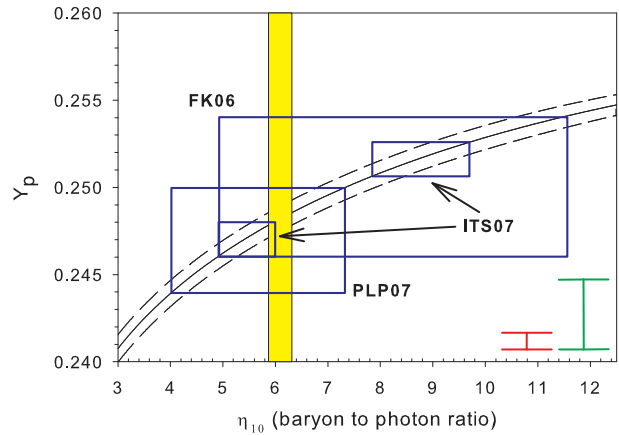}}
\caption{Primordial He mass fraction as a function of the baryon-to-photon ratio.  Several recent results are indicated.  See text for details.}
\label{fig:concordance}
\end{figure}

Figure~\ref{fig:concordance} plots the SBBN helium yields and errors (solid and dashed lines, respectively, Burles et al. 2001) as a function of $\eta_{10}$ ($=10^{10}~ \eta$).  The blue boxes are the one standard-deviation results of a few recent $Y_p$ estimates.  The higher and lower ITS07 boxes are using the Porter et al. (2005) and Benjamin, Skillman, \& Smits (1999) emissivities, respectively.  The vertical yellow bar depicts the $\eta_{10}$ (and associated error) derived from 3-Year WMAP observations (Spergel et al. 2007).  The red and green bars have heights equal to 2$\sigma$ ($\pm1\sigma$), where $\sigma$ are present results for optimistic and pessimistic extragalactic $\lambda 5876$ emissivities, respectively.  The green bar could be reduced with a full set of high quality recombination coefficients.  It could be quite difficult to convincingly narrow the optimistic results represented by the red bar, thus prohibiting a $dY/dZ$-based derivation of $\eta$ more accurate than those obtained by WMAP.  

PLP07 include a systematic error in $Y_p$ due to uncertainties in He~I emissivities.  Their value of $0.4\%$ and their $Y_p$ error are consistent with the present uncertainties in extragalactic $j(\lambda 5876)$.  The FK06 error in $Y_p$ is also consistent with our results.  Both the PLP07 and FK06 results are consistent with SBBN/WMAP.  The ITS07 results are inconsistent or marginally consistent with SBBN/WMAP, and their $Y_p$ errors are marginally consistent with our analysis.     

Correlated errors in pairs of He~I emissivitites could be important.
We checked ratios of $\lambda\lambda$4471, 5876, and 6678, in the optimistic extragalactic calculations and found little correlation; the line-ratio errors are roughly the quadrature sum of the individual line errors.

\begin{figure}
\protect\resizebox{\columnwidth}{!}
{\includegraphics{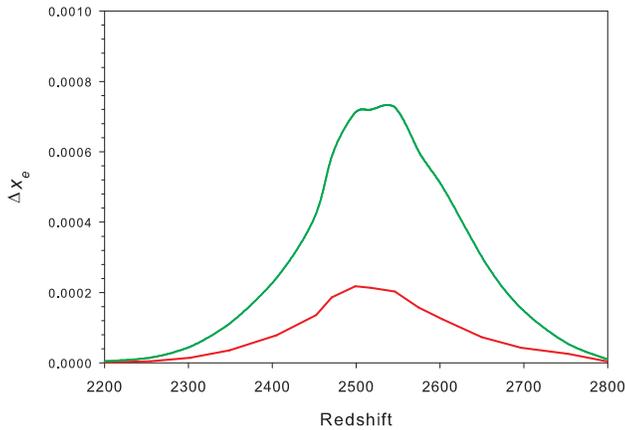}}
\caption{Fractional uncertainty in the electron fraction in the time-steady cosmological recombination model as a function of redshift.  Red and green curves are the optimistic and pessimistic cases, respectively.}
\label{fig:saha}
\end{figure}

Finally, we consider the cosmological case. In Figure~\ref{fig:saha}, the standard deviation in the electron fraction $x_e = n_e/n_p$ is plotted against redshift.  The pessimistic errors indicated by the green line reach $0.07\%$.  The optimistic errors illustrated by the red line are roughly a factor of 3 smaller.  Both are comparable to many of the effects considered by Switzer \& Hirata (2008).  A fuller \textit{non}-time-steady analysis will be considered elsewhere (Porter et al., in preparation).  

We thank Harry Nussbaumer for his helpful comments.  RLP and GJF thank the NSF (AST 0607028) and NASA (NNG05GD81G) for support.

\bsp

\label{lastpage}
\clearpage
\end{document}